\documentstyle[prc,aps,epsf,epsfig,amssymb,amsbsy,amstext,amsfonts]{revtex}

\begin{document}

\begin{center}
{\Large{\bf Two pion photoproduction on nucleons and nuclei in the $\rho$ and
$\sigma$ regions
  }}

\vspace{0.3cm}

\end{center}

\vspace{1cm}

\begin{center}
{\large{E. Oset, L. Roca, M.J. Vicente Vacas and J.C. Nacher}}
\end{center}

\begin{center}
{\small{ \it Departamento de F\'{\i}sica Te\'orica and IFIC, \\
Centro Mixto Universidad de Valencia-CSIC, \\
Ap. Correos 22085, E-46071 Valencia, Spain}}

\end{center}

\vspace{1cm}

\begin{abstract}
  In this talk we report on two physical processes, the photoproduction of
  $\rho$ mesons in nuclei and the $\pi^0 \pi^0$ photoproduction in nuclei 
at low energies.  In the first case the aim is to observe experimentally the
theoretically predicted changes on the $\rho $ properties in the nuclear medium.
In the second case one wishes to investigate the modifications of the $\pi \pi$
interaction in the nuclear medium in the region of the $\sigma$ meson. In
the first case we observe that it is quite difficult to see large effects due 
to the
$\rho$ modification in the medium because the detection of the $\rho $ is done
through the two pions which are produced in the $\Delta$ region and 
are largely distorted in the medium. In the second case, when the I=0 part of 
the $\pi \pi$ interaction is substituted by the
in medium  $\pi \pi$ amplitude, we observe a very large
shift of strength in the invariant mass distribution to small values of the mass
in $^{12}C$ and $^{208}Pb$ with respect to the distribution of the elementary
reaction. This spectacular shift appears to be corroborated by recent
experiments at Mainz reported in this same Workshop.  

\end{abstract}

\section{Index}

  In this talk I shall report on recent work concerning the photoproduction of
two pions in nuclei. The structure of the talk is as follows:\\ \\
II.\ \ \ Introduction\\
III.\ \ Model for $\gamma N \to \pi \pi N$\\
IV. \ \  $\rho$ photoproduction\\
V.\ \ \ \ $\rho$ photoproduction in nuclei\\
VI.\ \ \ $(\gamma,2\pi)$ in the $\sigma$ region\\
VII.\ \ $\pi \pi $ interaction in Unitary Chiral Perturbation Theory ($U\chi PT$)\\
VIII. The $\sigma$ as a $\pi \pi $ scattering resonance\\
IX.\ \ \ $\pi \pi $ interaction in the nuclear medium\\
X.\ \ \ \ $(\gamma,2\pi)$ in nuclei in the $\sigma$ region\\


\section{Introduction}
  The study of meson modification in a nuclear medium is a challenging problem
which attracts permanent theoretical and experimental attention. Even more
challenging is the modification of the meson meson scattering amplitudes inside
the medium. Pioneering work in this direction was done in \cite{Schuck:1988jn}
where the modification of the scattering amplitude led to the appearance of
peaks below the two pion threshold which were suggested as possible Cooper pairs
of two pions. In related work this phenomenon has been interpreted as a drop of
the $\sigma$ mass in the nuclear medium \cite{Hatsuda:1999kd} in a way that 
could be interpreted as
partial restoration of chiral symmetry. The observation of an accumulation of 
strength at low invariant
mass of two pions  in the pion induced $\pi^+\pi^-$  production 
\cite{Bonutti:1996ij,Bonutti:2000bv},
compared to the one in deuterium \cite{Bonutti:1998zw}, gave support to that
idea. More recently, in pion induced $\pi^0\pi^0$ production 
\cite{Starostin:2000cb}, the peaks seen in
$\pi^+\pi^-$ production in \cite{Bonutti:1996ij,Bonutti:2000bv} are not observed
but there is also accumulation of strength at lower invariant masses compared to
the free case. The most conclusive evidence could, however, come from photon induced 
two pion production experiments which are reported in the present
Meeting \cite{metag}.

 The theoretical work of \cite{Schuck:1988jn} stimulated many other works 
 where more detailed calculations were done 
\cite{Chanfray:1991fm,Mull:1992gt,Aouissat:1993uh}, qualitatively confirming the
findings of \cite{Schuck:1988jn}. More refined calculations imposing chiral
constraints in the $\pi \pi $ amplitudes \cite{Rapp:1996ir,Aouissat:1995sx} 
softened the response somewhat but still there was an enhancement of strength at
low invariant masses close to threshold compared to that in  free 
$\pi \pi $ scattering.

    The advent of $\chi PT$ as an approach to the QCD dynamics at low energies
\cite{Gasser:1984yg,Gasser:1985gg} has permitted to have a new look into the
problem. Yet, with all its success at low energies, $\chi PT$ has a limited
range of applicability and being perturbative in nature does not generate the
poles in the t-matrix. An important step forward, extending the predictive power
of $\chi PT$ to higher energies, has been done recently by reordering
the chiral expansion and imposing unitarity in coupled channels.  First steps
in this direction were done in \cite{Kaiser:1995eg} for the meson baryon 
interaction using a combination of the Lippmann Schwinger equation and input
from chiral Lagrangians, and in \cite{Oller:1997ti} in the meson meson
interaction using the Bethe Salpeter equation and just the lowest order
Lagrangian. Further refinements along this line have been done in 
\cite{Nieves:1999hp,Nieves:2000bx}. Another line of progress has been the use of
the Inverse Amplitude Method in coupled channels which allows one to
reproduce all the data of meson meson scattering up to 1.2 GeV
\cite{Oller:1998ng,Oller:1999hw,Guerrero:1999ei,GomezNicola:2001as}. Similarly,
another unitarization method \cite{Oller:1999zr} has been developed by means 
of the  N/D method,
together with the explicit exchange of genuine resonances which according to the
work of \cite{Ecker:1989te} account for
the second order Lagrangian of \cite{Gasser:1984yg,Gasser:1985gg}.  All these
methods, which are now known as unitarized chiral perturbation theory ($U\chi
PT$), give rise basically to the same results and reproduce very well the data
on meson meson scattering up to 1.2 GeV, leading to poles in the t-matrix for 
the different resonances appearing up to that energy.  Among these resonances,
the scalar ones, $\sigma(500)$, $f_0(980)$, $a_0(980)$, are generated
dynamically, meaning that with the use of the strong interaction provided by the
lowest order meson meson chiral Lagrangian, together with the unitarity
constraints which generate multiple scattering of the mesons, leads automatically
to these resonances without the need to introduce them explicitly in the
formalism. This is not the case for the vector mesons which require the
explicit inclusion in the N/D method of \cite{Oller:1999zr} or the implicit
inclusion through the use of the higher order Lagrangians in 
\cite{Oller:1998ng,Oller:1999hw}.
   The $\sigma$ meson which is of relevance to the present work is thus a
$\pi \pi$   scattering  resonance, and not a genuine QCD state which would
survive in the large $N_c$ limit, something already suggested in 
\cite{gasulf,ulfi}.  Hence the study of the medium modification of the $\pi \pi$
interaction around the $\sigma$ region and below with the $U\chi PT$ approach
seems most appropriate.  This work was carried out in \cite{Chiang:1998di} and
came to reconfirm the findings of the previous works where minimal chiral  
constraints were used \cite{Rapp:1996ir,Aouissat:1995sx}. Yet, a detailed
description of the pion induced two pion production in nuclei 
\cite{VicenteVacas:1999xx}, using the
information of \cite{Chiang:1998di}, does not lead to the peaks found in the
experiment \cite{Bonutti:1996ij,Bonutti:2000bv}. These results contrast with
those in \cite{Rapp:1999fx} where some approximations about the effective
density met by the pions were done, but due to a strong pion absorption the
effective densities are smaller than used in \cite{Rapp:1999fx},
as found in \cite{VicenteVacas:1999xx}.

  The $\rho$ renormalization in the medium has also been the subject of intense
debate and we address the reader to the review article \cite{Rapp:2000ej} for a
discussion of the different approaches and results.  For the purpose of the
present work it is sufficient to recall that the most recent works, including
\cite{Cabrera:2000dx} give a moderate shift of the $\rho$ mass to higher
energies and the width is increased by about 50 percent at normal nuclear matter
density.  In heavy ion reactions, where bigger densities can be reached,
experiments using dilepton detection show a qualitative evidence that the width
of the $\rho$ is increased although the results could also be interpreted in
terms of a dropping mass. 

  In the present paper we will try to investigate whether the predicted medium
modifications of the $\rho$ could be observed in some other reaction, concretely
in $\rho$ photoproduction in nuclei. Recent advances in the two pion
photoproduction have shown evidence of $\rho$ production 
\cite{Langgartner:2001sg} even at photon 
energies around 800 MeV, where only the tail
of the resonance is seen. The $\rho$ comes from the decay of
the D$_{13}$(1520)
resonance which is excited for photons around this energy. A recent theoretical
study of two pion photoproduction \cite{Nacher:2001eq}, improving the early 
results of 
\cite{GomezTejedor:1994bq,GomezTejedor:1996pe}, includes the mechanisms of $\rho$
production and also effects of the $\Delta(1700)$ excitation and leads to 
invariant mass distributions compatible with
the experimental ones found in \cite{Langgartner:2001sg}. Of course the $\rho$
is better seen at higher energies where the peak can be seen clearly in the
invariant mass of the two pions \cite{aachen}. So we have studied the
photoproduction at these higher energies, but not too high as to make the $\rho$
be produced with such large momenta in the nucleus that it decays outside the
nucleus, in spite of its large width, in which case we would only see the free
$\rho$ decay. 
 
\section{Model for $\gamma N \to \pi \pi N$}

\begin{figure}[ht!]
\centerline{\protect\hbox{\psfig{file=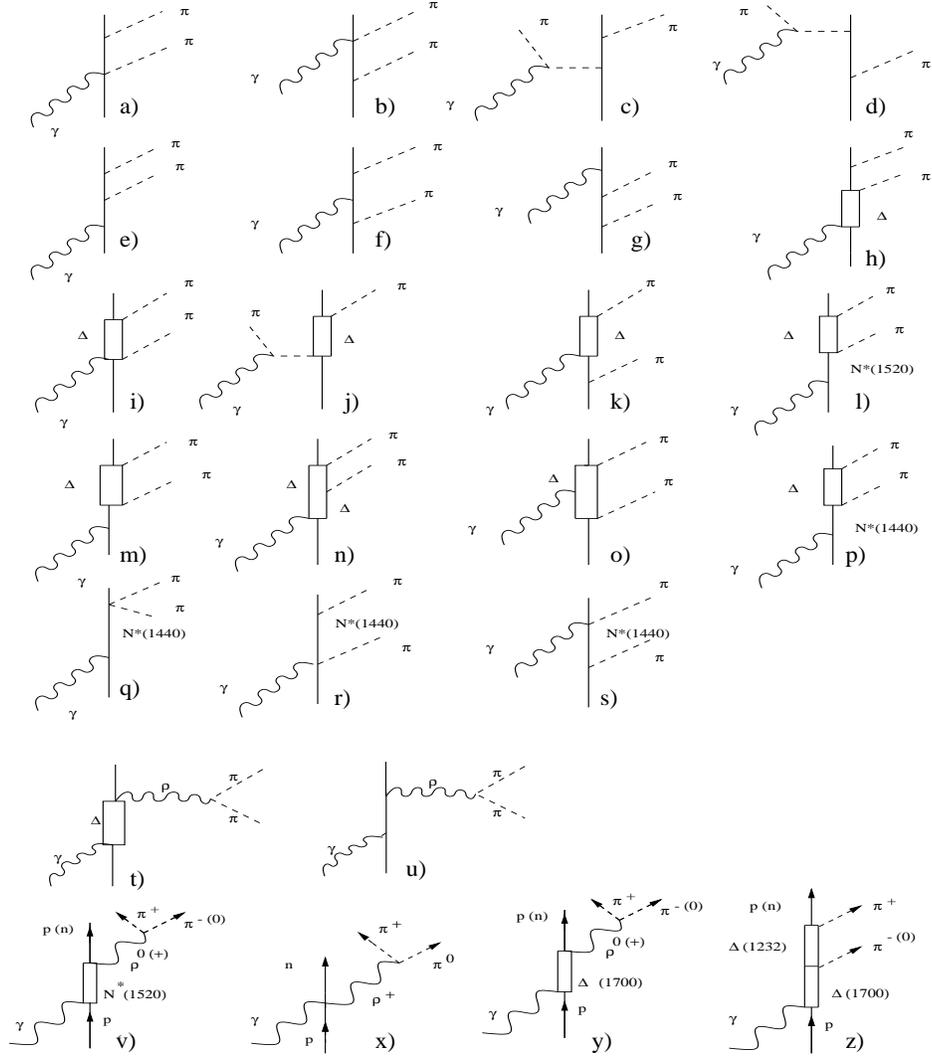,height=15.0cm,width=13.5cm,angle=0}}}
\caption{\small{Feynman diagrams for $\gamma N \to \pi \pi N$}}
\label{fig:diagrams}
\end{figure}

  As commented in the introduction we shall be using the updated model for
$\gamma N \to \pi \pi N$ of ref. \cite{Nacher:2001eq}. The model is meant to
cover energies up to 800 MeV, Mainz energies, where most of the experiments are
being performed. It contains tree level diagrams containing the coupling of
photons and pions to nucleons and resonances. The resonances included are the
$\Delta(1232)$, $N^*(1440)$, $N^*(1520)$ and $\Delta(1700)$. In addition two
pions can come out  as a $\rho$ meson from the decay of the $N^*(1520)$ and 
$\Delta(1700)$ resonances and from a $\gamma NN \rho$ contact term. The model is
depicted in figure \ref{fig:diagrams}.

\begin{figure}[h] \vspace{-0.5cm}
\centerline{\protect\hbox{\psfig{file=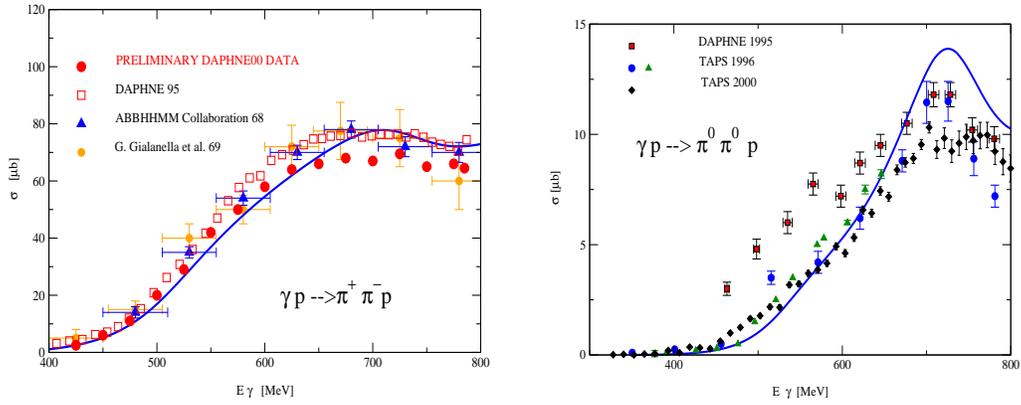,height=7cm,width=15cm,angle=0}}}
\caption{\small{Total cross section for $\gamma p \to \pi^+ \pi^- p$} and 
$\gamma p \to \pi^0 \pi^0 p$}
\label{fig:pi+pi-00}
\end{figure}

\begin{figure}[h]
\centerline{\protect\hbox{\psfig{file=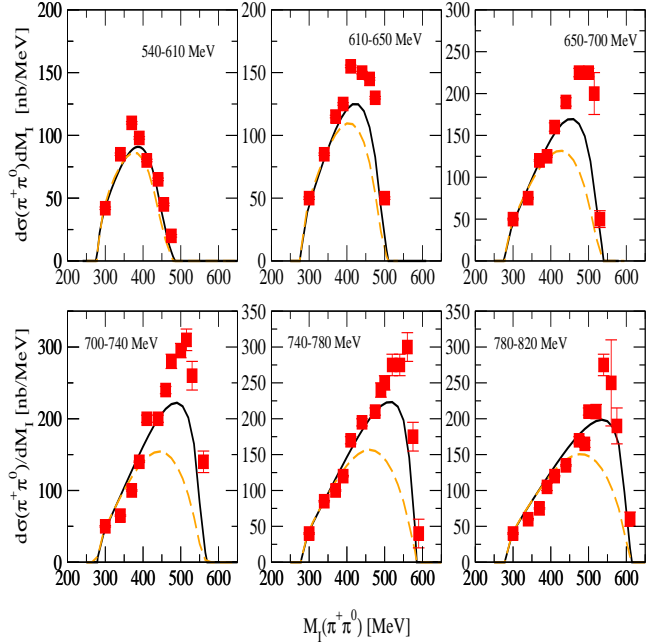,height=10cm,width=10.5cm,angle=-90}}}
\caption{\small{Two pion invariant mass distribution for $\gamma p \to \pi^+ \pi^0 p$. 
Continuous line: results with $\rho$ meson and $\Delta(1700)$ terms. Dashed line: 
results of the model without those contributions. 
Experimental data from ref. \protect\cite{Langgartner:2001sg} }}
\label{fig:pi+pi0}
\end{figure}

Results for the cross sections and invariant mass distributions can be seen in 
\cite{Nacher:2001eq}. We show in fig.\ref{fig:pi+pi-00} cross sections for $\pi^+\pi^-$ 
and $\pi^0\pi^0$ production on the proton together with experimental results
from several experiments. In fig. \ref{fig:pi+pi0} we show the invariant mass distribution of
the two pions in the case of $\pi^+\pi^0$ production on the proton. As one can
see there the effect of the $\rho$ production is important in shifting
strength to higher invariant masses in agreement with experiment. The model is
sufficiently accurate to be used to study the photoproduction of two pions in
nuclei.  We shall do that for two cases, the two pion production in a $\rho$
resonating state and the two pion production at low energies in the region of
the $\sigma$ meson.

\section{$\rho$ photoproduction}

\begin{figure}[ht!]\vspace{-0.5cm}
\centerline{\protect\hbox{\psfig{file=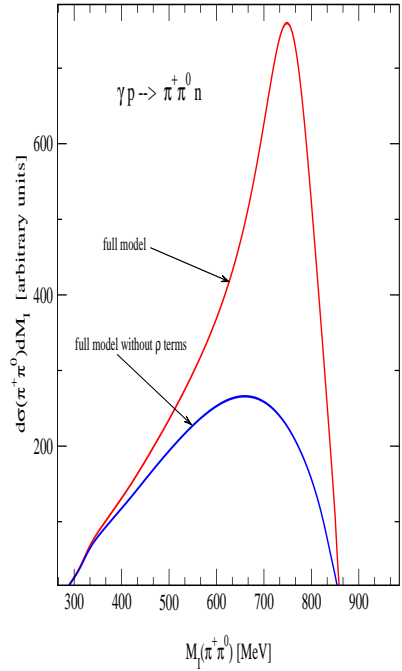,height=8cm,width=11.0cm,angle=-90}}}
\caption{\small{Prediction for free two pions invariant mass
distribution at $E_\gamma= 1250 MeV$}}
\label{fig:rho-libre}
\end{figure}

   In fig. \ref{fig:pi+pi0} we have seen how the $\rho$ production manifests itself in the
two pion invariant mass distribution. Since our aim is to explore the
possibility of finding medium effects through $\rho$ photoproduction, we have 
used the model for higher energies where there is sufficient phase space so
that the $\rho$ is seen as a peak in the invariant mass distribution. We are
aware that the model so far is only meant to work in the Mainz region but we
find that the shape obtained is in good agreement with the experiment    
\cite{aachen} although the absolute cross section is larger. Lack of unitarity
corrections from $\rho N$ final state interaction could be responsible for this 
larger strength, which however would not modify the shape of the  mass
distribution. Hence, since we simply want to compare the results for
photoproduction in nuclei and on the nucleon the extrapolation of the model can
serve very well the purpose. With this caveat, in fig.
\ref{fig:rho-libre} we show the results for
the $\gamma p \to \pi^+ \pi^0 n$ reaction at an energy $E_\gamma= 1250 MeV$. 

As we can see,
 the shape of the $\rho$ meson is clearly seen in the invariant mass
distribution. The peak reflects both the position and the width of the free
$\rho$ meson since the strength of the $\rho$ excitation is quite big compared 
to the background. 

\section{$\rho$ photoproduction in nuclei}

\begin{figure}[ht!]\vspace{-0.5cm}
\centerline{\protect\hbox{\psfig{file=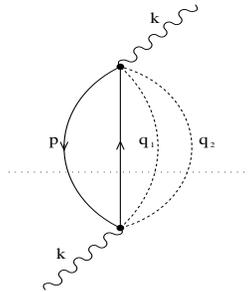,height=4cm,width=3.5cm,angle=0}}}
\caption{\small{Photon selfenergy to calculate the $\gamma N \to \pi \pi N$ 
cross section in nuclear matter}}
\label{fig:self}
\end{figure}

 In order to obtain the nuclear photoproduction cross section we evaluate the
photon selfenergy in nuclear matter of density $\rho$ due to the diagram
depicted in fig. \ref{fig:self}.

 The imaginary part of this diagram is obtained when the
intermediate states (a particle-hole and two pions) are placed on shell in the
integrations over the momenta of the intermediate states. The nuclear cross 
section is then given by \cite{Carrasco:1992vq}:

\begin{equation}\label{eq:sigma-a}
\sigma=-\frac{1}{k}\int d^3\vec{r}\ Im\Pi(k,\rho (r))
\end{equation}

where $\Pi(k,\rho (r))$ is the photon selfenergy and $k$ the photon momentum.
Equation (\ref{eq:sigma-a}) is making implicit use of the local density approximation, since the
photon selfenergy is evaluated at a fixed density and then this density is
replaced by the local nuclear density at the point $\vec{r}$ in the integral.

 The photon selfenergy corresponding to the diagram of fig. \ref{fig:self} is given by
 
 \begin{eqnarray} \label{eq:sigma1}
-i\Pi(k)=-\int \frac{d^4p}{(2\pi)^4} \int \frac{d^4q_1}{(2\pi)^4}
\int\frac{d^4q_2}{(2\pi)^4}\sum_{s_i,s_f}
\overline{\sum_{pol}}(-i)T(-i)T^{*}
\\ \nonumber & &
\hspace{-7.5cm}
 \cdot\frac{in(\vec{p})}{p^{0}-E(\vec{p})-i\eta}\,\,
\frac{i[1-n(\vec{k}+\vec{p}-\vec{q_1}-\vec{q_2})]}{k^{0}+p^{0}-q_{1}^{0}-q_{2}^{0}
-E(\vec{k}+\vec{p}-\vec{q_1}-\vec{q_2})+i\eta}
\\ \nonumber & &
\hspace{-7.5cm}
\cdot\frac{i}{q_{1}^{02}-\vec{q_{1}}^2-\mu^{2}+i\eta}
\,\,\frac{i}{q_{2}^{02}-\vec{q_{2}}^2-\mu^{2}+i\eta}
\end{eqnarray}
 
 and thus, performing the integration of the energy variables and substituting
 in eq. \ref{eq:sigma-a}, we find for the cross section
 
 \begin{eqnarray} \label{eq:sigma2}
\sigma=\frac{\pi}{k}\int d^3\vec{r}\int\frac{d^{3}\vec{p}}{(2\pi)^3}
\int\frac{d^{3}\vec{q_1}}{(2\pi)^3} 
\int\frac{d^{3}\vec{q_2}}{(2\pi)^3}\frac{1}{2\omega(\vec{q_1})}\frac{1}{2\omega(\vec{q_2})}
\\ \nonumber & &
\hspace{-7.5cm}
 \cdot\sum_{s_i,s_f}\overline{\sum_{pol}}\mid T\mid^{2}n(\vec{p})[1-n(\vec{k}+\vec{p}
 -\vec{q_1}-\vec{q_2})]
\\ \nonumber & &
\hspace{-7.5cm}
\cdot\delta(k^{0}+E(\vec{p})-\omega(\vec{q_1})-\omega(\vec{q_2})-E(\vec{k}
+\vec{p}-\vec{q_1}-\vec{q_2}))
\end{eqnarray}

 Eq. \ref{eq:sigma2} incorporates explicitly the Pauli blocking factor and assumes that
 there is no distortion of the particles.  This is actually not the case. The
 photon is not distorted but the pions can be absorbed in their way out of the
 nucleus from the point $\vec{r}$ of production which is the variable of
 integration in eq. \ref{eq:sigma2}. In order to take into account the distortion of the
 pions we make an eikonal approximation and remove from the pion flux those
 pions which undergo absorption, which indeed disappear, and also those which
 undergo quasielastic collisions because, even if they do not disappear, they
 change the momentum and energy of the pions in a considerable amount such 
 that the $\rho$ meson
 shape in the mass distribution will be lost and the events will go into the
 background. The eikonal factors that we consider are given by eq. \ref{eq:eikonal} and 
 the $\pi N$ cross sections, used for the distortion due to quasielastic steps,
 are taken from experiment, while the $C^{abs}$ coefficients are calculated
 theoretically from studies of the $\Delta$ selfenergy in a nuclear medium
 \cite{Oset:1987re} and tested against experimental pion absorption in nuclei in
 \cite{Salcedo:1988md}.
 
 \begin{eqnarray}\hspace{-1.1cm}\label{eq:eikonal}
F_i(\vec{r},\vec{q_i})=exp\left[-\int_{\vec{r}}^{\infty}dl_i\Big\{[\sigma_{\pi_i p}
\rho_p(\vec{r_i})+\sigma_{\pi_i n}\rho_n(\vec{r_i})]+C_i^{abs.(2)}\rho^{2}(\vec{r_i})
+C_i^{abs.(3)}\rho^{3}(\vec{r_i})\Big\}\right]
\\ \nonumber & & \vspace{0.3cm}\hspace{-13.2cm}
\vec{r_i}=\vec{r}+l_i \ \vec{q_i}/\mid \vec{q_i}\mid
\\ \nonumber & & \vspace{-0.5cm}\hspace{-13.2cm}
\sigma_{\pi_iN}=\textrm{quasielastic $\pi$-Nucleon cross section}
\\ \nonumber & &\vspace{-0.5cm}\hspace{-13.2cm}
C_i^{abs.}=\textrm{absorption coefficients by 2 and 3 body mechanisms}
\end{eqnarray}
 
  The pions coming from $\rho$ decay are mostly in the
 $\Delta$ resonance region where the studies of \cite{Salcedo:1988md} are done.
 For the few events where pions are produced with higher energies we use the
 parameterization of the absorption coefficients given in \cite{Oset:1990zj}.
 The superindex (2) and (3) in the absorption coefficients indicate absorption
 on two or three nucleons, a topic thoroughly investigated experimentally in the
 meson factories \cite{Weyer:1990ye}.
 
\begin{figure}[ht!]\vspace{-0.4cm}
\centerline{\protect\hbox{\psfig{file=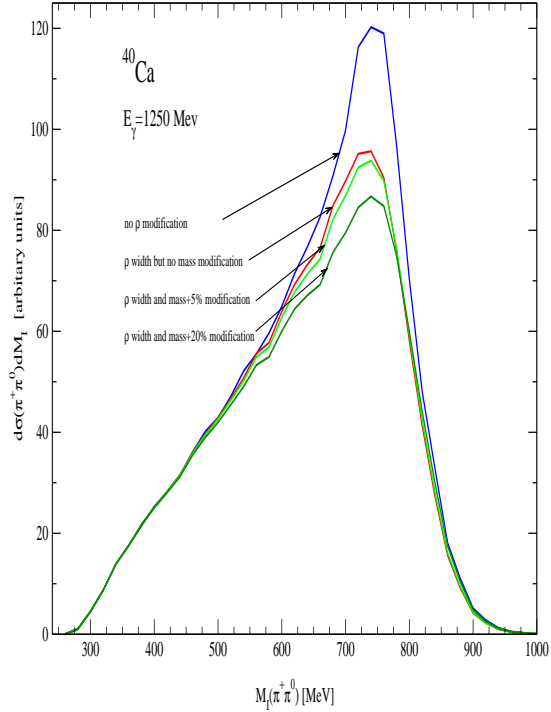,height=8cm,width=10cm,angle=-90}}}
\caption{\small{Two pion invariant mass distribution for $\pi^+\pi^0$ photoproduction on
$^{40}Ca$}}
\label{fig:rho-1250}
\end{figure}

    The results obtained for the invariant mass distribution in the nucleus of
 $^{40}Ca$ for $E_\gamma=1250 MeV$  are shown in fig. \ref{fig:rho-1250}.

  In the figure we show
 the results obtained using the free amplitude for $\gamma N \to \pi \pi N$ and
 those obtained by changing the width in the $\rho$ meson propagator by
\begin{equation}
\Gamma_{\rho}=\Gamma_{\rho free}(1+\frac{\rho}{\rho_0})
\end{equation}

This corresponds to doubling the $\rho$ width at nuclear matter density, which
seems to be about double from the predicted in the latest theoretical
approaches \cite{Rapp:2000ej},\cite{Cabrera:2000dx}. Even then, the results
obtained are rather moderate. The changes in the width induce about a 20 percent
decrease in the peak of the cross section and a corresponding increase of the
width in about the same amount. The effects found in the cross section are
much smaller than those assumed in the change of the in medium $\rho$ width.
This disappointing result must be seen in the fact that the pions produced are
mostly in the region of the $\Delta$ resonance where the probability that the
pions undergo  collisions or absorption is largest. There is also another
factor that goes against finding large effects of $\rho$ modification.
Unfortunately for this reaction, the $\rho$ is produced with a finite momentum
and then before it decays it travels a certain distance inside the nucleus
approaching the surface where the density dependent width resembles more the
free width. We take this into account using in the integration the $\rho$
width at the point reached by the $\rho$ after its lifetime in the medium, which 
we take at the 
intermediate point between the point of production and the point of decay.

   We have also checked the effects due to a hypothetical mass change of the
$\rho$ and the results found, even for large changes in the mass like 150 MeV
at normal nuclear matter density, are very moderate as can be seen in the
figure.

   This finite traveling inside the nucleus is a handicap for observation of
the medium effects, which is avoided in the dilepton production processes where
one tries to maximize effects by looking to dileptons back to back, which would
imply production of the $\rho$ at rest.  One could try to minimize this effect
by putting cuts in the $\rho$ momentum and concentrating only in events where
the $\rho$ is produced with a small momentum, which would correspond to events
where the $\rho$ is produced backwards in the $\gamma N$ CM frame and one takes
further advantage of the Fermi motion. This procedure was successfully used in 
\cite{Oset:2001na} in order to find medium effects in $\phi$ photoproduction.
 Yet, one was looking there for slow kaons which were not so
drastically distorted as the pions from $\rho$ decay. We have done similar tests
here but the results are still very similar to those shown in fig. \ref{fig:rho-1250},
 only with smaller cross sections.

  In view of the present results and discussion, the conclusion seems to be that
trying to see medium effects of the $\rho$ from $\rho$ photoproduction in nuclei
is quite a difficult task.  The fact that the $\rho$ is produced with a finite
momentum, and most important, that the pions are produced in the resonance region
where the distortion of the pions is very large, has as a consequence that the
process is very peripheral and tests very small densities where the medium
modifications are small.

\section{$(\gamma,2\pi)$ in the $\sigma$ region}

In section III we could see that the agreement with experiment of the 
$\gamma p \to \pi^0 \pi^0 p$ reaction is overall good but in the region of the
$\sigma$, $E_\gamma$ around 400 to 500 MeV, our cross section is smaller than
experiment.  Actually, this is an interesting point, since as shown in 
\cite{Bernard:1994ds}, the consideration of chiral loops at threshold increased
appreciably the $\pi^0\pi^0$ production rate.  Here we will find similar
results in the region of the $\sigma$, but in order to show this we need to
introduce the elements of $\pi \pi $ scattering from the perspective of 
$U \chi PT$, which we review briefly in the next section.

\section{$\pi \pi $ interaction in Unitary Chiral Perturbation Theory }

\begin{figure}[ht!]
\centerline{\protect\hbox{\psfig{file=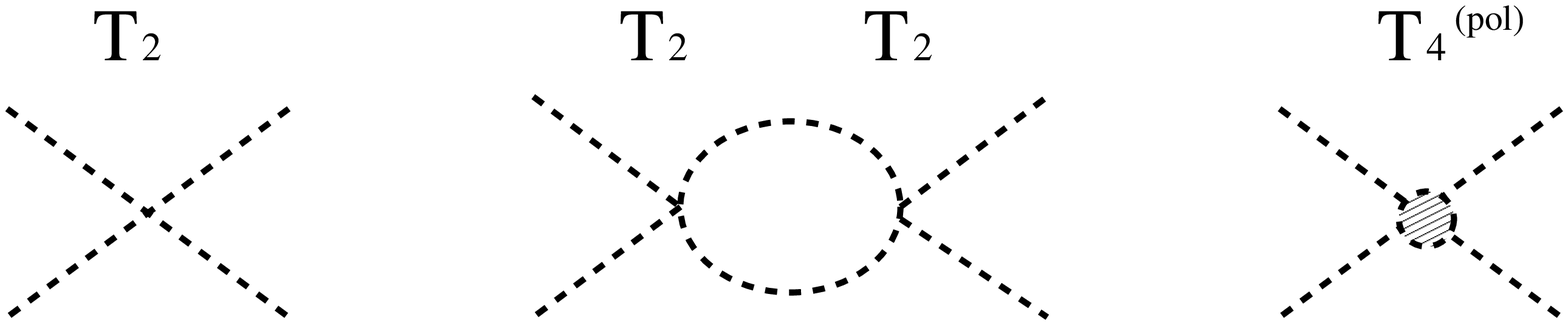,height=2.2cm,width=9cm,angle=0}}}
\caption{\small{Contributions to de T matrix up to $O(p^4)$}}
\label{fig:T4}
\end{figure}

The starting point in this discussion is the chiral Lagrangian 
\cite{Gasser:1985gg}

\begin{equation}\label{eq:L2}
L_{2}=\frac{f^{2}}{4}<\partial_{\mu} U^{\dag}\partial^{\mu}U+M(U+U^{\dag})>
\end{equation}

where the symbol $< >$ stands for the trace of the SU(3) matrices involved in the
Lagrangian, $f$ is the pion decay constant, M the mass matrix 
$diag(m_\pi^2,m_\pi^2,2m_K^2-m_\pi^2)$ and $U$ is the standard SU(3) matrix
involving the meson fields \cite{Gasser:1985gg,Meissner:1993ah,Pich:1995bw}. In
addition there would be a second order Lagrangian involving the $L_i$
coefficients of Gasser and Leutwyler \cite{Gasser:1985gg}.
  In $\chi PT$, omitting crossed loops, the lowest order Lagrangian provides the
$O(p^2)$ contribution to the meson meson scattering matrix, $T_2$, while the 
$O(p^4)$ contribution comes from the loop in fig. \ref{fig:T4} with the two vertices
obtained from the lowest order Lagrangian, plus the polynomial contribution from
the second order Lagrangian, $T_4(pol)$.

 The $T$ matrix to order $O(p^4)$ is then given by
 
\begin{equation}
T= T_2 +T_4
\end{equation}

The unitary extensions described in the introduction rely upon the
implementation of exact unitarity in coupled channels

\begin{equation}
Im T = T \sigma T^*
\end{equation}

\begin{equation}
\sigma \equiv \sigma_{ll}=-\frac{k_{l}}{8\pi\sqrt{s}}
\theta \left(s-(m_{1l}+m_{2l})^{2} \right)
\end{equation}

where $\sigma$ is a diagonal matrix( in the space of the coupled channels) which
accounts for the phase space of the intermediate two mesons which we consider in
our approach ($\pi \pi$ and $K \bar{K}$ in our case).  In order to connect with 
$\chi PT$ we realize that $\sigma$ is the imaginary part of the loop function of
two mesons in fig. \ref{fig:T4}, $G_{ll}$

\begin{equation} 
G_{ll}=i\int\frac{d^{4}q}{(2\pi)^{4}}\frac{1}{q^{2}-m_{1l}^{2}+i\epsilon}
\frac{1}{(P-q)^{2}-m_{2l}^{2}+i\epsilon}
\end{equation}

 \begin{equation}
 \sigma_{ll}=Im G_{ll}
 \end{equation}
 
  Hence we can write
\begin{equation}
Im T= T ImG T^*\to ImG= -Im T^{-1}  
\end{equation}

\begin{equation}
T^{-1}=Re T^{-1}- i Im G$ $\to$ $T=(Re T^{-1}-i Im G)^{-1}
\end{equation}  
 
 Although the different unitary approaches discussed at the beginning are
 technically different, the essence of all of them is that an expansion in
 powers of $O(p^2)$ is done for $Re T^{-1}$, not $T$, and by virtue of this one
 succeeds in providing:
\begin{enumerate}
\item Faster convergence of the expansion
\item Larger convergence radius in the energy variable
\item The meson resonances up to 1.2 GeV
\end{enumerate}

In the IAM method one gets now expanding $ReT^{-1}$ up to $O(p^4)$
\begin{equation}\label{eq:13}
T=T_2(T_2-T_4)^{-1}T_2
\end{equation}

Now it is interesting to note that $T_4$ is fixed in chiral perturbation theory
but neither the loop contribution not the $T_4(pol)$ are defined since they both
depend on the regularization scheme applied to renormalize the loops, only their
sum is fixed. Now assume that we are able to choose a renormalization scheme
( a cut off in \cite{Oller:1997ti}) such that the $T_4(pol)$ contribution is
minimized. In such case $T_4$ is just $T_2GT_2$ and then eq. \ref{eq:13} becomes 
\begin{equation}
T=(1-T_2G)^{-1}T_2$ $\to$ $T-T_2GT=T_2
\end{equation}

which is nothing but the Bethe Salpeter equation with $T_2$ as a kernel
(potential)
\begin{equation}
T=T_2+T_2GT
\end{equation}

This procedure justifies why in \cite{Oller:1997ti} one could get good results
for the scalar sector in terms of only the lowest order Lagrangian. 
In particular one obtained the scalar resonances $\sigma(500)$, $f_0(980)$
$a_0(980)$ and $\kappa(900)$. The results
obtained with this procedure are practically identical to those found in
\cite{Oller:1998ng,Oller:1999hw} with the IAM and explicit use of the second
order Lagrangian (see \cite{Oller:2000ag} for updated results). However, the
same procedure could not be used to generate the vector mesons and explicit use
of the second order Lagrangians was needed in the IAM, or explicit exchange of
vector meson resonances had to be assumed in \cite{Oller:1999zr}.  This is why
we associate these vector mesons to genuine QCD states while claim that the
scalar mesons are dynamically generated.

\section{The $\sigma$ as a $\pi \pi $ scattering resonance}
The $\sigma(500)$ is generated in this approach with the Bethe Salpeter equation
and the lowest order chiral amplitude used as kernel and comes out as a broad
resonance which is found as a pole in the second Riemann sheet and has a mass 
$m_{\sigma}=470 MeV$ and a width around $\Gamma_{\sigma}= 400 MeV$. These
results would be in good agreement with the recent experimental analysis of
 \cite{Aitala:2001xu}, with $m_\sigma=478^{+24}_{-23}\pm 17$ MeV and 
$\Gamma_{\sigma}=324^{+42}_{-40}\pm 21$.

\section{The $\pi \pi $ interaction in the nuclear medium}
The $\pi \pi$ interaction in a nuclear medium in the $L=I=0$
channel ($\sigma$ channel) has stimulated much theoretical work
lately.
 As commented in the Introduction, it was realized that the attractive 
 P-wave interaction of the pions
with the nucleus led to a shift of strength of the $\pi \pi$ system
to low energies and eventually produced a bound state of the two
pions around $2 m_\pi  -  10$  MeV \cite{Schuck:1988jn}. This state
would behave like a $\pi \pi$
Cooper pair in the medium, with repercussions in several
observable
magnitudes in nuclear reactions \cite{Schuck:1988jn}. The possibility
that such
effects could have already been observed in some unexpected
enhancement
in the ($\pi, 2 \pi$) reaction in nuclei \cite{2} was also
noticed there.
More recent experiments where the enhancement is seen in the
$\pi^+ \pi^-$
channel but not in the $\pi^+ \pi^+$ channel \cite{Bonutti:1996ij} have added
more
attraction to that conjecture (see also the talk of Starostin \cite{starostin}
in this Workshop).

\begin{figure}[ht!]\vspace{-0.4cm}
\centerline{\protect\hbox{\psfig{file=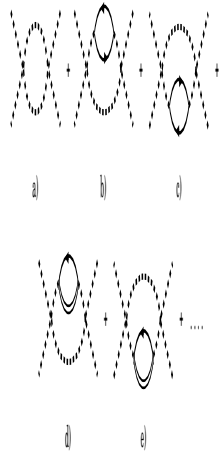,height=3cm,width=6.75cm,angle=-90}}}
\caption{\small{Terms appearing in the scattering matrix allowing the
pions to excite $ph$ and $\Delta h$ components }}
\label{fig:pipi1}
\end{figure}

\begin{figure}[ht!]\vspace{-2cm}
\centerline{\protect\hbox{\psfig{file=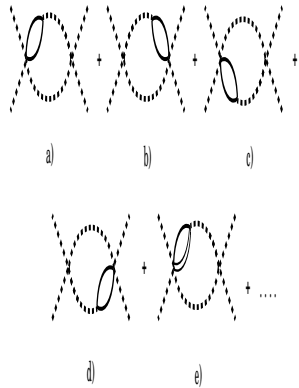,height=4.5cm,width=8.1cm,angle=-90}}}
\caption{\small{ Terms of the $\pi \pi$ scattering series in the nuclear
medium related
to three meson baryon contact terms}}
\label{fig:pipi2}
\end{figure}

\begin{figure}[ht!]\vspace{-0.4cm}
\centerline{\protect\hbox{\psfig{file=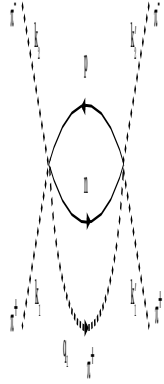,height=2.5cm,width=5.8cm,angle=-90}}}
\caption{\small{ Diagram involving the three meson baryon contact terms}
of fig. 
\ref{fig:pipi2} 
in each of the vertices }
\label{fig:pipi3}
\end{figure}

\begin{figure}[ht!]
\vspace{0.5cm}
\centerline{\protect\hbox{\psfig{file=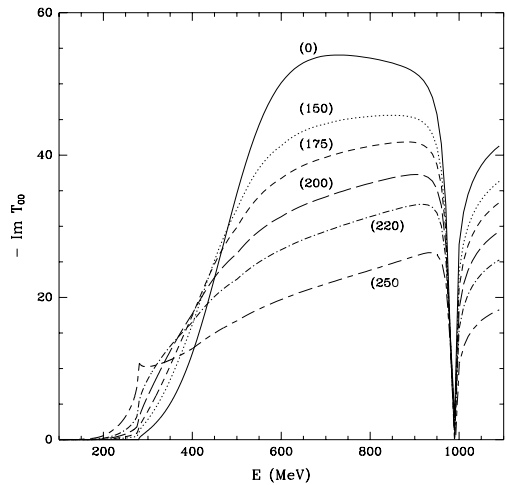,height=6cm,width=6.5cm,angle=0}}}
\caption{\small{Im $T_{22}$  for $\pi \pi \rightarrow \pi \pi $}
scattering in $J=I=0$
 $(T_{00}$ in the figure) in the
nuclear medium for different values of $k_F$ versus the CM energy
of the 
pion pair. The labels correspond to the values of $k_F$ in MeV. }
\label{fig:pipi4}
\end{figure}

  The advent of the chiral unitary methods has added new interest in the
  subject and has allowed one to focus on the implications  of the chiral
  constraints which had been known to be relevant in this kind of studies
  \cite{Rapp:1996ir}. In \cite{Chiang:1998di} the $\pi\pi$ interaction in a nuclear medium was
  studied following the lines of \cite{Oller:1997ti}, renormalizing the pion 
  propagators in the medium and introducing vertex corrections for
  consistency. These vertex corrections play an interesting role since, as proved
  in \cite{Chanfray:1999nn}, they exactly cancel the off shell contribution of
  the two pion loops with renormalized pions. The diagrams considered are 
  depicted in figs. \ref{fig:pipi1},\ref{fig:pipi2},\ref{fig:pipi3}.  The results
  for the imaginary part of the $\pi\pi$ amplitude in L=I=0 are shown in fig.
  \ref{fig:pipi4}.
    One can appreciate that there is an accumulation of strength in the region
    of small invariant masses of the two pion system, with qualitative results
    similar to those found in \cite{Schuck:1988jn}, but we do not get poles in that
    energy region.  The accumulation of strength at these small invariant
    masses could raise hopes that the enhancement of strength at small invariant
    masses found in the $(\pi,2\pi)$ reactions in nuclei in \cite{Bonutti:1996ij} could be
    explained. However, according to a recent study \cite{VicenteVacas:1999xx}, the 
    small nuclear densities involved in this reaction,
    which is rather peripheral, make the changes found in \cite{Chiang:1998di}
    insufficient to explain the experimental data.
    
      The work of \cite{Chiang:1998di} includes only the $\pi\pi$ channel, since one is
 only concerned about the low energy region. This work has been generalized 
 to coupled channels in \cite{manolonew} in order to  make predictions for 
 the modification of the $f_0(980)$ and $a_0(980)$ resonances in a nuclear 
 medium.  One finds there that  both  resonances become wider in the
 medium as the nuclear density increases, with the $a_0(980)$
 eventually melting into a background for densities close to normal nuclear
 matter density. The $f_0(980)$  resonance, which in the free space is
 narrower than the $a_0(980)$, still would keep its identity at these high
 densities but with a width as large as 100 MeV or more.  How to produce these
 resonances in a nucleus in order to check the predictions of these studies is a
 present experimental challenge.

\section{$(\gamma,2\pi)$ in nuclei in the $\sigma$ region}
  Unitarity is lost when one works with tree level Feynman diagrams.
Yet, this is only partly true because in the Feynman diagrams one is usually
including all possible resonances where one takes the width into account and 
then the most
important part of the amplitude is included in a unitarized form. This would be
the case when one has a delta in the final state, where the $\pi N$ channel
would be  unitarized. However, the final state of two pions is not
unitarized. Once more this is only partly true because we explicitly account for
$\rho$ production, which means that the unitarization of the two pion state in 
the L=1, I=1 channel is taken into account.  But the unitarization in the
L=0,I=0 is not done and we do not explicitly include a $\sigma$ production. As
mentioned in section III, perturbative unitarity is taken into account in 
\cite{Bernard:1994ds} by means of loops at energies close to threshold. Here we
wish to account for the final state interaction in L=0 for energies of the
photon about 400-460 MeV, where Mainz experiments are being performed
\cite{metag}. For that purpose we realize that at these energies the pions come
mostly in S-wave. We actually carry out a projection of the amplitudes in this
channel and find indeed that this is the case. However, we could have isospin
I=0 or I=2. Since the interaction of pions at these energies in I=2 is very
weak we implement the effects of final state interaction only in the I=0, for
what we have to separate the pion production amplitude in two parts. This is
done easily. We first write the isospin decomposition of the $\pi^0 \pi^0$ state
as 

\begin{eqnarray}\label{eq:T00}
|\pi^0(1)\pi^0(2)>=
\underbrace{\frac{1}{3}|\pi^0(1)\pi^0(2)+\pi^+(1)\pi^-(2)+\pi^-(1)\pi^+(2)>}
_{\textrm{I=0 part}}
\\ \nonumber & & \hspace{-7.1cm} 
\underbrace{-\frac{1}{3}|\pi^0(1)\pi^0(2)+\pi^+(1)\pi^-(2)+\pi^-(1)\pi^+(2)>
+|\pi^0(1)\pi^0(2)>}_{\textrm{I=2 part}}
\end{eqnarray}

Hence the $\gamma N \to N \pi^0 \pi^0$ amplitude can be decomposed in two parts:
the one that has as final state the combination of I=0, which requires to make
the linear combination of amplitudes requested by the linear combination of 
$\pi \pi$ states in the I=0 term, first term of the RHS of eq. \ref{eq:T00}, and the I=2 
combination, last two terms of eq. \ref{eq:T00}, which we
leave as it is at the tree level.

\begin{figure}[ht!]
\vspace{-0.4cm}
\centerline{\protect\hbox{\psfig{file=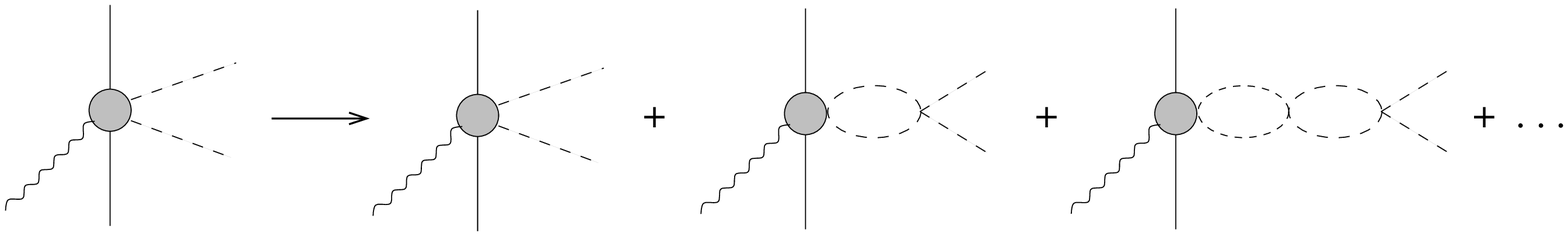,height=2.5cm,width=15.0cm,angle=0}}}
\caption{\small{Diagrammatic series for pion final state interaction in I=0}}
\label{fig:Tgammapipi}
\end{figure}

 To account for the pion final state interaction
in I=0 we sum up the diagrammatic series in fig \ref{fig:Tgammapipi}, 
which analytically can be expressed as

\begin{equation}
T_{(\gamma,\pi^0\pi^0)}(I_{\pi\pi}=0)\to T_{(\gamma,\pi^0\pi^0)}(I_{\pi\pi}=0)
\left(1+G_{\pi\pi}t_{\pi\pi}^{I=0}(M_I)\right)
\end{equation}

where $G_{\pi\pi}$ is the G function for the loop function of the two pions, used
before in the Bethe Salpeter equation, and $t_{\pi\pi}^{I=0}$ is the
$\pi\pi$ scattering matrix in isospin I=0.  By making use of the findings in
\cite{Oller:1997ti} and \cite{Oller:1999zr}, which tell us that in the two pion 
loop function the T matrix can be factorized on shell, we assume this to be the
case also for the tree level two pion production amplitude and this leads
immediately to the equation written above.  We have checked that this is a good
approximation at low energies, but at the energies where we are working it has
to be improved a bit and corrections of the order of 25 percent arise from
these corrections.  The main idea is that the dominant terms as we go to higher
energies  is the $\Delta$ Kroll Ruderman term, see diagram (i) of Fig (1). This
term is present in the $\gamma N \to \pi^+ \pi^- N$ amplitude which appears
in the I=0 combination. The sum of the two amplitudes leading to 
$\pi^+(1) \pi^-(2)$ and $\pi^-(1) \pi^+(2)$ in the final state leads to a term
which contains the factor $\vec{S}(\vec{p}_{\pi_1}+\vec{p}_{\pi_2})\vec{S}^{\dag}
\vec{\epsilon}(\gamma)$
and furthermore there are three propagators in the loop, the two pion
propagators and the $\Delta$ propagator. It is important to keep the propagator
in the loop function because since one is approaching the situation in which the
delta is placed on shell in the tree diagrams, small  variations of the
energy and momentum variables of the $\Delta$ in the loop lead to differences
with respect to the factorization of the tree level amplitude.  This loop
function is done following the same steps as the vertex function evaluated in
\cite{Oset:2000gn}, where the $\sigma$ exchange potential was calculated by
allowing two pions interact in s-wave and using again the same chiral techniques
exposed here.

We also include two extra baryon form factors in the loop, as done in 
\cite{Oset:2000gn}, to account for the $\pi N \Delta$ vertex correction. After
this is done, we find it technically useful in order to account for the more elaborate
loop corrections discussed above, to still apply the factorization of the
$(\gamma, 2 \pi)$ tree level amplitude  but with a slightly modified form
factor included in the $G_{\pi\pi}$ loop function. This procedure is quite
accurate numerically and prevents the numerical task from blowing up when we
perform the calculations in nuclei.

  There is also a small technical detail. One of the terms in our approach
contains the Roper excitation and its posterior decay into two pions in S-wave.
The term is given by the Lagrangian

\begin{equation}
{\mathcal{L}}_{N^*\pi\pi}=-C_{1}^{*}m_{\pi}^2\overline{\Psi}_N\vec{\phi}^2\Psi_{N^*}
\end{equation}

where the constant $C_1^*$ is fitted to the partial decay width. Since this
term is already projected in the L=0, I=0 channel, we substitute 
\begin{equation}
C_1^*\to C_1^*\frac{1}{1+G(\overline{M_I})t_{\pi\pi}^{I=0}(\overline{M_I})}
\end{equation}
where $\bar{M}_I$ is an average $\pi\pi$ invariant mass in the decay of the Roper
into a nucleon and two pions,
such that when the renormalization due to the interaction is done explicitly
in our approach in the region of energies where the Roper is excited, then we
obtain the empirical term that leads to the observed partial decay width.

   The cross section for the process is now given by the same equation
   \ref{eq:sigma2}, where now the distortion factors are
   
 \begin{eqnarray}\hspace{0cm}
F_i(\vec{r},\vec{q_i})=exp\left[\int_{\vec{r}}^{\infty}dl_i \frac{1}{q_i}Im
\Pi (\vec{r}_i) \right]
\\ \nonumber & &\vspace{0.4cm}\hspace{-5.5cm}
\vec{r_i}=\vec{r}+l_i \ \vec{q_i}/\mid \vec{q_i}\mid
\end{eqnarray}

where $\Pi$ is the pion selfenergy.  Here we are in the region of low energy 
pions and for the pion selfenergy we use the results of 
\cite{Nieves:1993ye,Nieves:1993ev}. This potential has the advantage to have
been tested against the different reaction cross sections, elastic, quasielastic
and absorption. The imaginary part of the potential is split into a part that
accounts for the probability of quasielastic collisions and another one which
accounts for the pion absorption probability. As we shall see, the probability
that there is loss of pion flux through pion absorption at low energies is 
 larger than that where 
there are quasielastic collisions. One of the reasons is the Pauli
blocking of the occupied states.

 On the other hand, and this is one of the important points of the work, when
we renormalize the I=0 amplitude to account for the pion pion final state
interaction we change $G(s)$ and $t_{\pi\pi}(I=0)$ by their corresponding results
in nuclear matter evaluated at the local density of the point $\vec{r}$ in the
integral of eq. \ref{eq:sigma2}.

We have also used the $\Delta$ selfenergy from \cite{Oset:1987re} to dress the
$\Delta$ propagator. In addition to the proper real part of the selfenergy in 
\cite{Oset:1987re} we add the effective contribution to the selfenergy 
$4/9(f^{*}/\mu)^2g'\rho$ coming from the iterated
$\Delta$h excitation driven by de Landau Migdal interaction \cite{Carrasco:1992vq} 
\cite{Oset:1987re}.

\begin{figure}[ht!]
\vspace{-0.4cm}
\centerline{\protect\hbox{\psfig{file=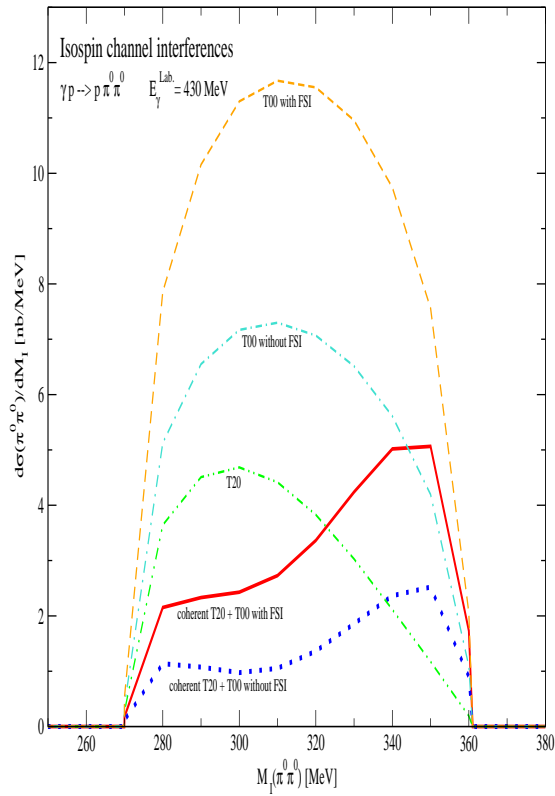,height=8cm,width=11.0cm,angle=-90}}}
\caption{\small{Contributions of the different isospin channels to the $2\pi$
 mass distribution.}}
\label{fig:interf}
\end{figure}

   With all this introduction we pass now to present the results.  In the first
place, in fig \ref{fig:interf} we can see the results of the invariant mass distribution of
the two pions for the $\gamma p \to \pi^0 \pi^0 p$ reaction.

In the figure 
we can see the contribution of the I=0 part alone,
the part of I=2 alone and the coherent sum of the two, both in the case when the
I=0 amplitude is renormalized and when it is not.  We can see that the
renormalization of the I=0 amplitude has important effects nearly doubling
the cross section.  When then we sum coherently the I=0 and I=2 amplitudes we
observe a curious shape of the distribution with a double hump, one at low
invariant masses and the other one at the high mass part of the spectrum.  This
shape is corroborated by experiment as seen in the presentation in this Workshop
by Volker Metag \cite{metag}. 

\begin{figure}[ht!]
\vspace{-0.4cm}
\centerline{\protect\hbox{\psfig{file=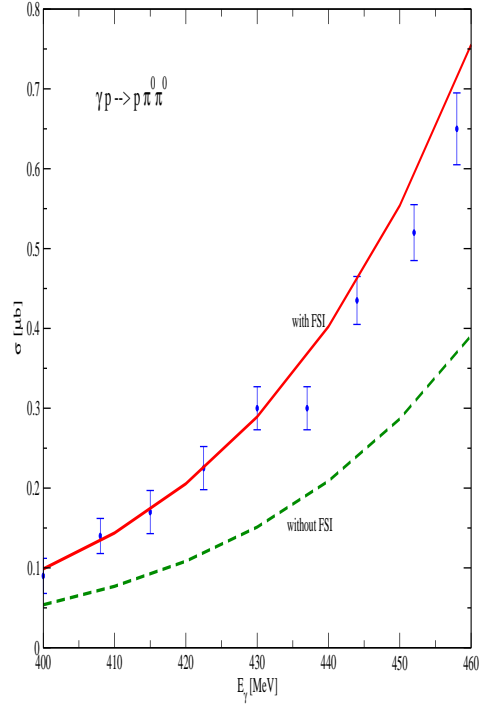,height=7cm,width=10.0cm,angle=-90}}}
\caption{\small{Total cross section for $\gamma p \to \pi^0 \pi^0 p$ with and without pion
final state interaction. Experimental data from ref. \protect\cite{Wolf:2000qt}.}}
\label{fig:X-free}
\end{figure}

  The integrated cross section compared with experiment can be seen in fig. \ref{fig:X-free}
where we can appreciate that the inclusion of final state interaction has lead
to an improvement of the cross section.

\begin{figure}[ht!]
\vspace{-0.3cm}
\centerline{\protect\hbox{\psfig{file=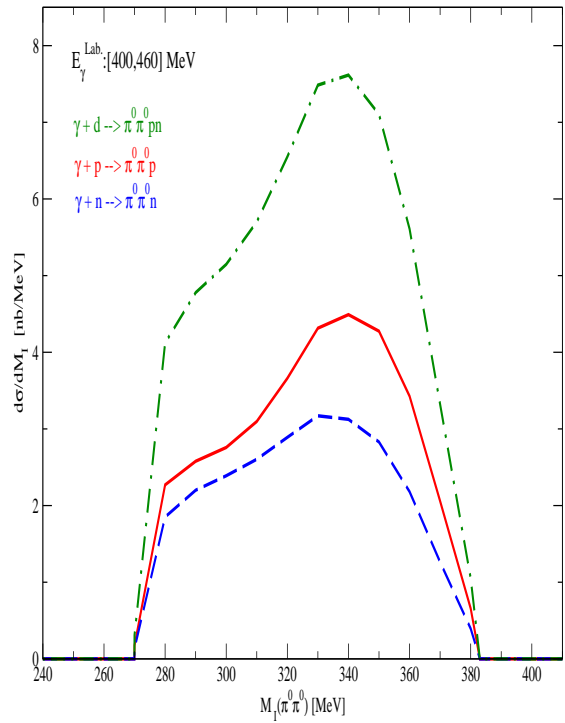,height=8.5cm,width=10.2cm,angle=-90}}}
\caption{\small{Two pion invariant mass distribution for $2\pi^0$ photoproduction on
proton, neutron and deuteron, (continuous, dashed and dashed-dotted line respectively).}}
\label{fig:inv-free}
\end{figure}

  In fig. \ref{fig:inv-free} we show the results of the invariant mass distribution for 
$\gamma p \to \pi^0 \pi^0 p$, $\gamma n \to \pi^0 \pi^0 n$ and
$\gamma d \to \pi^0 \pi^0 pn$, the latest one obtained as a simple sum of the
two other cross sections.

   Next we show the results of the invariant mass distribution in nuclei.
   
\begin{figure}[ht!]
\vspace{-0.3cm}
\centerline{\protect\hbox{\psfig{file=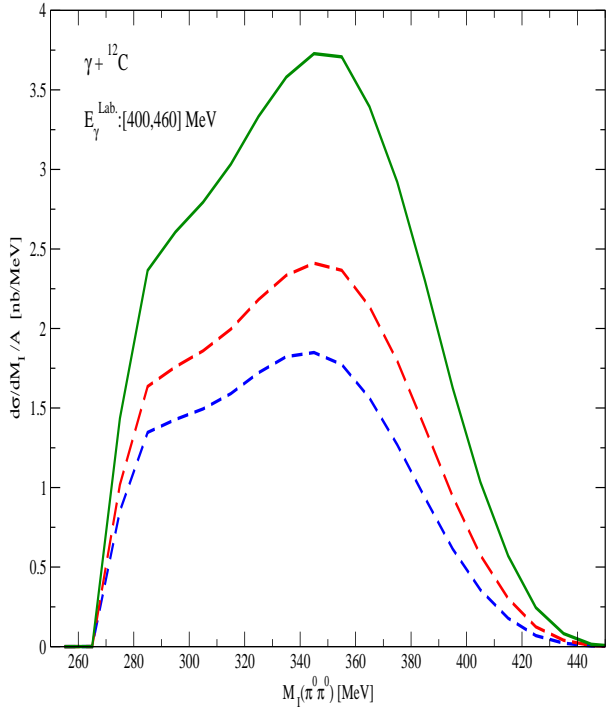,height=8.3cm,width=9.96cm,angle=-90}}}
\caption{\small{Two pion invariant mass distribution for $2\pi^0$ photoproduction 
in $^{12}C$. All three curves are calculated using the $2\pi$ final state interaction at
 0 density but they differ in the final pion distortion:
continuous line: without absorption nor quasielastic scattering. Long dashed line: only final
pions absorption. Short dashed line: final pions absorption and quasielastic scattering.}}
\label{fig:C12-rho0}
\end{figure}

\begin{figure}[ht!]
\vspace{-0.4cm}
\centerline{\protect\hbox{\psfig{file=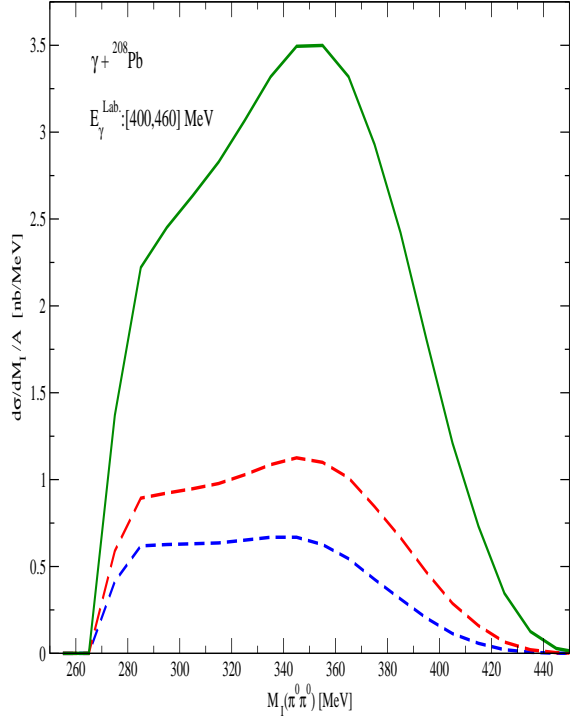,height=8.5cm,width=10.2cm,angle=-90}}}
\caption{\small{Same as fig. \ref{fig:C12-rho0} for  $^{208}Pb$.}}
\label{fig:Pb208-rho0}
\end{figure}

 First we show in fig. \ref{fig:C12-rho0} the results for $^{12}C$ without including 
 the pion pion final state interaction in nuclei,
  meaning we use the free $\gamma N \to \pi \pi N$
 amplitude including the final state interaction of the pions in free space.
 However, we include the pion distortion. We can see that there is a reduction
 of about 40 percent in the cross section due to pion absorption. If we remove
 the pions which have undergone a quasielastic scattering this would reduce the
 cross section in an extra 20 percent.  This effect is moderate. Part of these
 collisions would not change the charge of the pions, only their energy and
 momentum would be changed. In this case the two $\pi^0$ would still be there
 and their invariant mass would be somewhat changed. In other cases there could
 be change of charge and then we would not have two $\pi^0$ in the final state.
 However, this could also be compensated by having originally  the $\pi^+\pi^0$
  production
 followed by a collision of the $\pi^+$ with charge exchange. We shall show
 results eliminating only the pions absorbed but we can take the differences
 between the calculations removing only the pions absorbed, or the one where we
 remove all pions which are absorbed or undergo quasielastic collisions, as 
 indicative of the theoretical uncertainties.     
 The results for the case of removal of only the pions which are absorbed are
 obtained  by putting in the imaginary part of the pion selfenergy, $\Pi$,
 the part which comes from the absorption and omitting the one that comes from
 quasielastic, which have been separated in \cite{Nieves:1993ye}. In fig. 
 \ref{fig:Pb208-rho0} 
 we can see the same results in $^{208}Pb$, which are qualitatively similar but
 where the amount of pions absorbed is much larger.  
 
   Now comes the interesting part when the medium effects are included in the
pion pion final state interaction. 

\begin{figure}[ht!]
\vspace{-0.4cm}
\centerline{\protect\hbox{\psfig{file=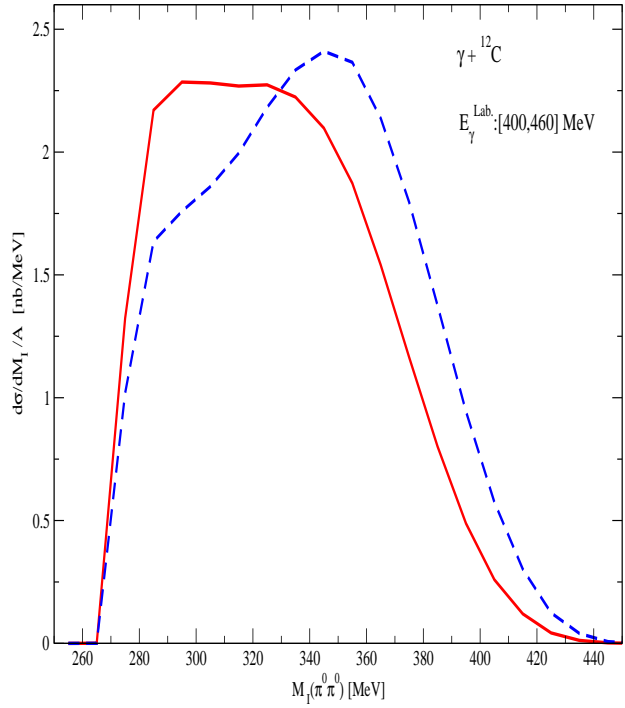,height=8.5cm,width=10cm,angle=-90}}}
\caption{\small{Two pion invariant mass distribution for $2\pi^0$ photoproduction 
in $^{12}C$. Continuous line: Using the in medium final $\pi\pi$ interaction. Dashed line: 
using the final $\pi\pi$ interaction in free space.}}
\label{fig:C12}
\end{figure} 

In fig. \ref{fig:C12} we can see the results for
$^{12}C$.  The only difference between the two curves has been the use of the in
medium $\pi \pi$ scattering and $G(s)$ function instead of the free ones.  As
one can see in the figure  there is an appreciable shift of strength to the low
invariant mass region due to the in medium  $\pi \pi$  interaction. The
accumulation of strength found in  all approaches close to the two pion
threshold has its manifestation in this impressive shift of strength.  This
shift is remarkably similar to the one shown by Metag \cite{metag} in this
Workshop. 

\begin{figure}[ht!]
\vspace{-0.4cm}
\centerline{\protect\hbox{\psfig{file=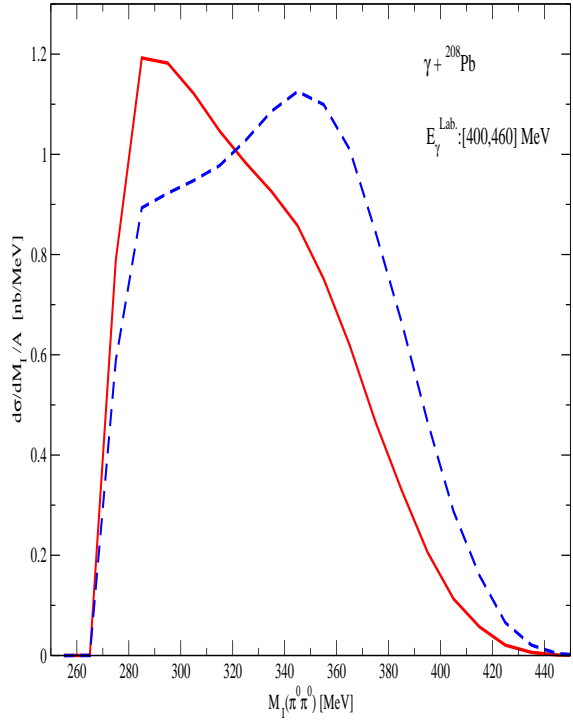,height=8.5cm,width=10.2cm,angle=-90}}}
\caption{\small{Same as fig. \ref{fig:C12} for $^{208}Pb$. }}
\label{fig:Pb208}
\end{figure}

  In fig. \ref{fig:Pb208} we can see the same results for $^{208}Pb$ which look
 qualitatively similar, but the shift to low invariant masses is further accused 
 in this case, as it also happens in the experiment \cite{metag}.  

   This is in our opinion the first clear manifestation of this modified pion
pion interaction in the medium, given the fact that there are problems in the
comprehension of the pion induced two pion production experiments and their
theoretical description.  The fact that the photons are not distorted has
certainly an advantage and allows one to see inner parts of the nucleus. On the
other hand there is an extra advantage because the I=0 amplitude for the case of
$\pi^0 \pi^0 $ production was quite large since it involves the $\pi^+ \pi^- $
production amplitudes which are much larger than the $\pi^0 \pi^0 $ one.

\section{conclusions}

  We touched here two  issues looking for medium modifications of the mesons.
  The first one was about the modifications of the $\rho$ meson.  Even when most
authors agree about the important medium modifications of the $\rho$ meson
properties, what we found here is that seeing these results in $\rho $ meson
photoproduction in nuclei is very difficult. The fact that the pions are
produced in the $\Delta $ resonance region distorts the pions considerably and
only regions of low densities close to the surface can be tested where the
medium effects are small.

  On the other hand we could see that $\pi^0 \pi^0$ photoproduction in nuclei 
  at low
energies is  an excellent reaction to see the medium effects of the pion pion
interaction close to the two pion threshold.  This increased strength  can be
rightly attributed to the moving of the $\sigma$ resonance at low energies in
the nuclear medium, and its decreased width, as suggested in 
\cite{Hatsuda:1999kd} and found recently
within the framework of $U \chi PT$ by M.J. Vicente Vacas and reported in this
Workshop \cite{sigmamedio}.

\subsection*{Acknowledgments}

One of us, L.R. acknowledges support from the Consejo Superior de
Investigaciones Cientificas.
This work is also
partly supported by DGICYT contract number BFM2000-1326, and the 
E.U. EURODAPHNE network contract no. ERBFMRX-CT98-0169.

\end{document}